# NEW CCD PHOTOMETRIC OBSERVATIONS OF W-UMA ECLIPSING BINARY NSV 5904 USING THE 1.88-m KOTTAMIA TELESCOPE, EGYPT

A. Essam[1], A. Naklawy[1], A. A. Haroon[1], M. A. Hamdy[1], G. B. Ali[1], H. A. Ismail[1], I. M. Selim[1], Y. A. Azzam[2], and F. I. Elnagahy[3]

**ABSTRACT:** *New BVR light curves of the eclipsing binary system NSV 5904 have been constructed based on CCD observations obtained using 1.88-m telescope of Kottamia observatory during the phase of telescope testing and adjusting its optical quality on May, 2009. New times of minima and epoch have been determined from these light curves. Using the Binary Maker 3.0 (BM3) package, a preliminary determination of the photometric orbital and physical parameters of NSV 5904 are given.*

## INTRODUCTION

NSV 5904 (DF CVn = TYCHO2 3021.2642.1; $\alpha 2000 = 12^h 43^m 37\overset{s}{.}24$; $\delta 2000 = +38°44'.15.''70$) was discovered as a suspected variable star by Weber (1963) during a photographic survey of selected areas in the northern hemisphere, on the basis of a few observations showing light changes. It was catalogued as NSV 5904 in the New Suspected Variable Stars (Kholopov 1982). Visual observations of this system have been obtained by some members of the amateur's European association GEOS. They were able to confirm the variability of the system as eclipsing binary of the EW sub-class. Also, they determined the first preliminary period, Vandenbroere (1999).

Photoelectric measurements were performed at the Jungfraujoch station on the basis of the collaboration between GEOS and Geneva Observatory (Vandenbroere et al. 2001); nine BV points were obtained during two nights on December 1998. A very small amplitude of the B-V colour index suggested the eclipsing nature of NSV 5904.

Vandenbroere et al. (2001) also obtained more than 200 unfiltered CCD

---
[1] *Astronomy Dept., National Research Institute of Astronomy and Geophysics (NRIAG)*
[2] *Computer Science Dept., College of Science in Azzulfi, Majmaah University, Saudi Arabia*
[3] *Computer Science Dept., Faculty of Computing and Information Technology, King Abdul Aziz University, Jeddah, Saudi Arabia*

images using a TI245C camera mounted on 0.2m Newtonian reflector. They derived four times of minimum light and determined the new ephemeris as follow:

Min I = HJD 2450571.219 + 0.326890 × E

On the basis of these results, the system had a designation as DF CVn. Acerbi et al. (2005) published the first V filter light curve based on CCD observations using Kodak KAF 401 chip mounted on a 0.2-m Schmidt-Cassegrain telescope for seven nights from JD 2453107 to JD 2453145. They used the WD method to analyze the light curves of DF CVn and they found out that the star DF CVn is a contact A-type W Ursae Majoris variable star, with a mass ratio q = 0.347 and the primary (deeper) minimum occurs when the larger, more massive star, is eclipsed by its smaller, less massive companion.

## PRESENT OBSERVATIONS

Photometric observations of the eclipsing binary system NSV 5904 have been obtained in B,V, and R (closely match to those of the standard Jonson system) wide pass-band filters through five nights, $15^{th}$, $16^{th}$, $17^{th}$, $18^{th}$, and $21^{st}$ of May 2009, using EEV CCD 42-40 camera with a format of 2048*2048 pixels, cooled by liquid nitrogen to -120 C° attached to the Newtonian focus of the 1.88-m Kottamia reflector telescope at Egypt. The B and R observations for the system NSV 5904 are presented for first time to this star.

Differential photometry was performed with respect to GSC 3021-2613 and GSC 3021-2617, as a comparison and check stars respectively, which were photometrical calibrated as standard stars. All times were corrected to HJD. The photometric calibration of the raw CCD images was performed according to the standard method as specified by the following formula:

REDUCED = [(raw) - (bias) - (master dark)] / (master flat)

The recent time of primary minimam in V filter derived by Dvorak (2009) and the period derived by Acerbi et al. (2005) are used to construct the following ephemeris:

Hel. JD. (Min.I) = 2454503.7561 + 0.3268956*E          (1)

This ephemeris is used to derive the normal points of light curves Δb, Δv, and Δr (Differential Instrumental Magnitude) on the nights 16, 17, and 18 of May 2009, which are illustrated in Figure 1 and listed in Table A1 of Appendix A.

___________________________________________________________

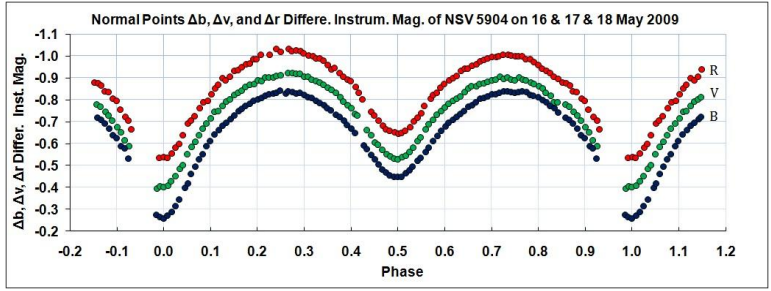

**Fig. 1: Normal points light curves Δb, Δv, and Δr (Differential Instrumental Magnitude) of NSV 5904 for three nights, 16$^{th}$, 17$^{th}$, and 18$^{th}$ of May 2009**

## EPOCHS OF PHOTOMETRIC MINIMA

New thirteen times of minima of NSV 5904 (four primaries and nine secondaries) were derived from the present photometry. The moments of these minima were calculated using the software package AVE (Barbera, 1996), that employs the method of kwee & Woerden (1956). Table 1 gives the following parameters: date of observation, the deduced times of minima (HJD), the probable error in the times of minima (P.E.), type of minima (Min.), the filter type (Filter), the O-C's residuals (O-C), and the number of integer cycle (E) according to the ephemeris given in equation 1, respectively.

No significant difference (within error quoted) was found between the times of minima of B, V, and R filters. This implies that, it is not a function of colour. Hence, their weighted means can be used for phasing the nightly light curve and period behaviour study.

**Table 1: Epochs Minimum Light of NSV 5904**

| Date | HJD=2454900+ | P.E. | Min. | Filter | O-C | E |
|---|---|---|---|---|---|---|
| 15 May 2009 | 67.460919 | 0.000262 | II | B | 0.010433 | 1418.5 |
| | 67.461088 | 0.000689 | II | V | 0.010950 | 1418.5 |
| | 67.461159 | 0.000546 | II | R | 0.011167 | 1418.5 |
| 16 May 2009 | 68.273472 | 0.000564 | I | B | -0.003902 | 1421.0 |
| | 68.274418 | 0.001380 | I | V | -0.001008 | 1421.0 |
| | 68.274975 | 0.000991 | I | R | 0.000696 | 1421.0 |
| | ٦٨.٤٣٧٤٣١ | 0.000093 | II | B | -0.002338 | 1421.5 |
| | 68.437351 | 0.000278 | II | V | -0.002583 | 1421.5 |
| | 68.437436 | 0.000360 | II | R | -0.002323 | 1421.5 |
| 17 May 2009 | 69.254716 | 0.000181 | I | B | -0.002198 | 1424.0 |
| 18 May 2009 | 70.398743 | 0.000142 | II | B | -0.002527 | 1427.5 |
| | 70.399130 | 0.000047 | II | V | -0.001343 | 1427.5 |
| | 70.399030 | 0.000144 | II | R | -0.001649 | 1427.5 |

## PHOTOMETRIC PERIOD DETERMINATION

The photometric orbital period of the system NSV 5904 is derived using the FORTRAN program "PERDC" written by our late colleague "Dr. A. El-Bassuny Alawy" in private communication. The program based on Ferraz method (Ferraz 1981), where Fourier series has been applied. Only the B observations achieved on 16th, 17th, and 18th of May were combined and analyzed. This is due to two reasons. First, both minima and maxima were observed; i.e. all characteristic phases are well represented. Second, the observations have been obtained in short time span (three consecutive nights). This may minimize errors resulted from light curve variation (a common phenomenon shown by most of late W UMa systems).

We have determined a period of 0.32682465 days, with correlation coefficient equal 0.875974 and amplitude of $0.432_m$, in normalized magnitude, which are in good agreement with the light curve amplitude observed. Using the mean time of primary minima of the second night (16th of May 2009) and the new period, we obtained the new ephemeris to phase our observations as follows:

$$\text{HJD (Min.I)} = 2454968.273472(\pm 0.000564) + 0.32682465*E \qquad (2)$$

Where E is the number of integer cycle.

We applied this ephemeris on three nights of observations (16th, 17th, and 18th of May 2009) to obtain the normalized light curves in B, V, and R filters in differential magnitude, as presented in Figures 2, 3, and 4 respectively.

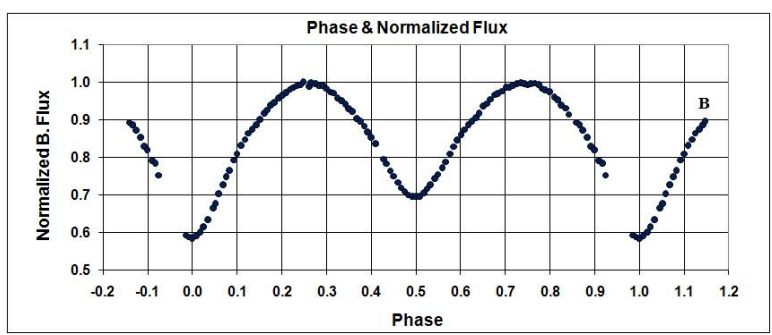

**Fig. 2: Normalized B light curve of NSV 5904 for three nights, 16th, 17th, and 18th of May 2009**

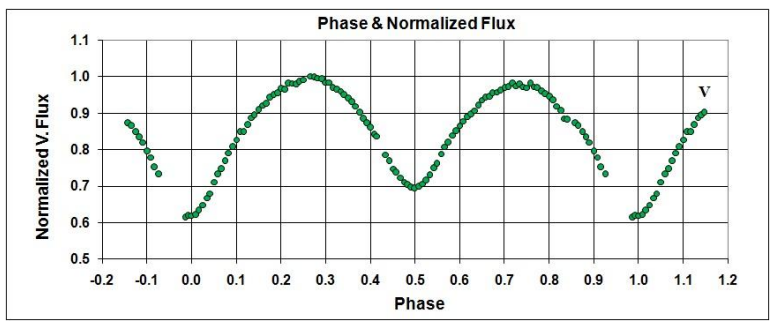

**Fig. 3: Normalized V light curve of NSV 5904 for three nights, 16th, 17th, and 18th of May 2009**

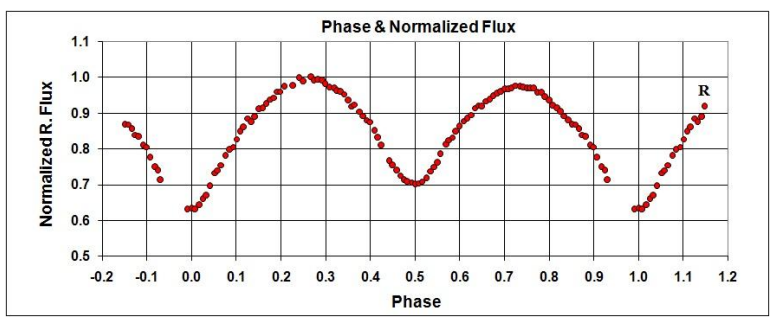

**Fig. 4: Normalized R light curve of NSV 5904 for three nights, 16th, 17th, and 18th of May 2009**

## PRELIMINARY PHOTOMETRIC DATA ANALYSIS

Binary Maker 3, (BM3 hereafter), Bradstreet (2005), (see for software description) was used to construct a theoretical model for the system NSV 5904. The software requires normalized phase-flux data (See Figures 2, 3, and 4) to perform the modeling for each filter (B, V, and R). The accurate modeling of any eclipsing binary system requires both light and radial velocity curves. In our case and due to the lacking of the radial velocity curves for the system under study, we construct a model based on the previously published models with the same basic shape of the light curve of this system.

The various parameters are then 'tweaked' until the theoretical light curve perfectly matches the actual light curve. The parameters for the known objects are obtained from the library supplied in BM3 CD or from the Catalogue and Atlas of Eclipsing Binaries (CALEB) website at http://caleb.eastern.edu.

Our theoretical models construction is based on a models trend similar to that of the observed light curves.

The observed light curves of the system NSV 5904 reflect the occurring of a partial eclipsing event, which enables us to estimate the degree of inclination. Also, we used different values of temperature, until we get the best fit between theoretical and observed light curves.

The variations between the primary and secondary maxim for each V and R light curves mean that some light variation was evident from one side of the system to the other (O'Connell-effect) (O'Connell, 1951). By adding a cool spot on the surface of the larger component, we could treat this light variation.

The BM3 output parameters are listed in Table (2) for B, V, and R filters. The preliminary fitting between the normalized observed light curves and the normalized theoretical light curves which were generated by BM3 program are shown in Figures 5, 6, and 7 for B, V, and R filter respectively.

Table 2: BM3 parameters for NSV 5904

| **Main Parameters** | **B** | **V** | **R** |
|---|---|---|---|
| *Mass Ratio (q)=$M_1/M_2$* | 0.3665 | 0.3665 | 0.3665 |
| *Surface potential ($\Omega_1 = \Omega_2$)* | 2.5465 | 2.5465 | 2.5465 |
| *Fillout* | 27 % | 27 % | 27 % |
| *Surface Temp. $T_1$* | 5330 °k | 5330 °k | 5330 °k |
| *Surface Temp. $T_2$* | 4850 °k | 4980 °k | 4980 °k |
| *Inclination (i)* | 75.0 | 75.0 | 74.8 |
| *Lumnusty1=$L_1/(L_1+L_2)$* | 0.8189 | 0.7724 | 0.7646 |
| *Lumnusty2=$L_2/(L_1+L_2)$* | 0.1811 | 0.2276 | 0.2354 |
| *Main Radius1* | 0.4853 | 0.4853 | 0.4853 |
| *Main Radius2* | 0.3126 | 0.3126 | 0.3126 |
| **Spot** | | | |
| *Co-latitude* | | 90° | 90° |
| *Longitude* | | 15° | 15° |
| *Radius* | | 15% | 15% |
| *Temp factor* | | 0.51 | 0.51 |
| *L. C. Residual* | 0.0109 | 0.1129 | 0.0108 |

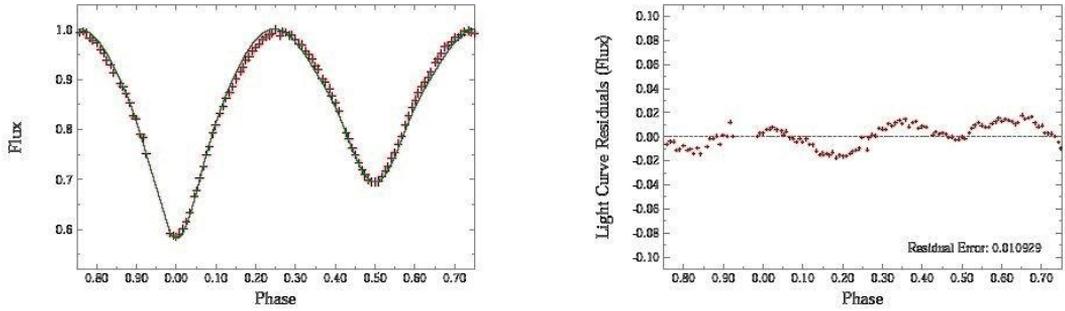

**Fig. 5: The normalized B light curve for NSV 5904 (crosses) overlaid with the theoretical light curve generated by BM3 (solid line) and LC Residual in right corner**

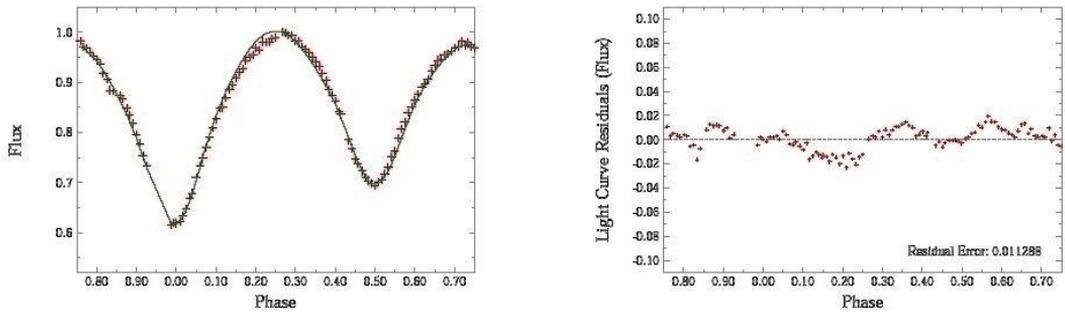

**Fig. 6: The normalized V light curve for NSV 5904 (crosses) overlaid with the theoretical light curve generated by BM3 (solid line) and LC Residual in right corner**

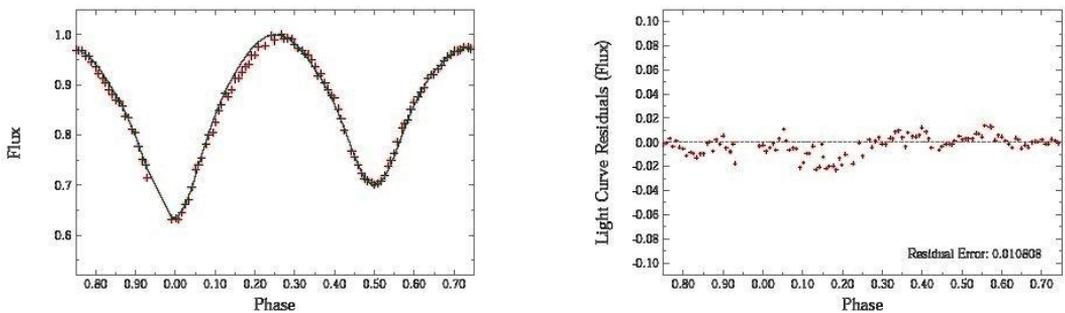

**Fig. 7: The normalized R light curve for NSV 5904 (crosses) overlaid with the theoretical light curve generated by BM3 (solid line) and LC Residual in right corner**

## CONCLUSIONS AND FUTURE REMARKS

The B and R observed light curves of the system NSV 5904 are presented for the first time.

The NSV 5904 system exhibits slightly light variation from one side to the other (O'Connell-effect) (See O'Connell, 1951) in V and R filters as seen in the Figures 3 and 4. This effect can be well explained by starspot hypothesis on the components of W UMa systems that posses high surface activity due to their rapid rotation and convective envelopes.

New thirteen times of minima of NSV 5904 (four primaries and nine secondaries) were derived from the present photometry. The new linear ephemeris was determined as:

$$HJD (Min.I) = 2454968.273472(\pm 0.000564) + 0.32682465*E$$

The V and B preliminary solutions provide almost similar values for orbital inclination and the components' physical parameters (mass ratio, surface potential and surface temperature). These parameters are, in principle, agreed with those derived by others (e.g. Acerbi et al. 2005).

The solutions reveal that NSV 5904 is an over contact binary system by 27%, a shallow common convective envelope that may be responsible for its high activity.

An attempt to model the system NSV 5904 was made using the BM3, but systematic deviations between observational points and fitting curve are always found. A more sophisticated method (like WD method) combined with multi wavelength observations should allow a better description for the system NSV 5904 and may be our future works that we intent to do.

# APPENDIX A
### Table A1: Normal Points of light curves Δb, Δv, and Δr for the system NSV 5906

| No. | No. of Points | Phase | Δb | P.E. | No. of Points | Phase | Δv | P.E. | No. of Points | Phase | Δr | P.E. |
|---|---|---|---|---|---|---|---|---|---|---|---|---|
| 1 | 2 | -0.1416 | -0.7178 |  | 1 | -0.1425 | -0.7770 |  | 1 | -0.1486 | -0.8791 |  |
| 2 | 1 | -0.1341 | -0.7098 |  | 2 | -0.1350 | -0.7677 |  | 1 | -0.1409 | -0.8766 |  |
| 3 | 2 | -0.1266 | -0.6924 |  | 2 | -0.1250 | -0.7452 |  | 2 | -0.1334 | -0.8641 |  |
| 4 | 2 | -0.1166 | -0.6692 |  | 1 | -0.1175 | -0.7277 |  | 1 | -0.1260 | -0.8398 |  |
| 5 | 1 | -0.1090 | -0.6385 |  | 2 | -0.1099 | -0.7067 |  | 2 | -0.1184 | -0.8351 |  |
| 6 | 2 | -0.1015 | -0.6265 |  | 2 | -0.0999 | -0.6752 |  | 2 | -0.1083 | -0.8039 |  |
| 7 | 2 | -0.0907 | -0.5880 |  | 1 | -0.0915 | -0.6507 |  | 1 | -0.1008 | -0.7951 |  |
| 8 | 1 | -0.0833 | -0.5775 |  | 2 | -0.0842 | -0.6154 |  | 2 | -0.0929 | -0.7562 |  |
| 9 | 2 | -0.0758 | -0.5322 |  | 2 | -0.0742 | -0.5872 |  | 2 | -0.0826 | -0.7204 |  |
| 10 | 2 | -0.0153 | -0.2734 |  | 2 | -0.0137 | -0.3960 |  | 1 | -0.0751 | -0.7061 |  |
| 11 | 3 | -0.0079 | -0.2647 | 0.0043 | 2 | -0.0077 | -0.4034 |  | 1 | -0.0701 | -0.6667 |  |
| 12 | 3 | -0.0008 | -0.2593 | 0.0022 | 4 | -0.0002 | -0.4023 | 0.0072 | 4 | -0.0091 | -0.5341 | 0.0211 |
| 13 | 4 | 0.0082 | -0.2711 | 0.0034 | 3 | 0.0088 | -0.4090 | 0.0073 | 3 | 0.0004 | -0.5367 | 0.0121 |
| 14 | 3 | 0.0172 | -0.2896 | 0.0093 | 3 | 0.0159 | -0.4284 | 0.0075 | 3 | 0.0075 | -0.5341 | 0.0146 |
| 15 | 3 | 0.0243 | -0.3145 | 0.0168 | 4 | 0.0249 | -0.4515 | 0.0071 | 4 | 0.0165 | -0.5553 | 0.0143 |
| 16 | 4 | 0.0335 | -0.3462 | 0.0139 | 3 | 0.0341 | -0.4844 | 0.0147 | 3 | 0.0255 | -0.5829 | 0.0170 |
| 17 | 1 | 0.0457 | -0.3997 |  | 2 | 0.0403 | -0.5028 |  | 3 | 0.0326 | -0.5967 | 0.0226 |
| 18 | 3 | 0.0509 | -0.4194 | 0.0118 | 3 | 0.0499 | -0.5526 | 0.0195 | 3 | 0.0409 | -0.6396 | 0.0280 |
| 19 | 3 | 0.0584 | -0.4600 | 0.0187 | 4 | 0.0589 | -0.5861 | 0.0131 | 3 | 0.0515 | -0.6925 | 0.0348 |
| 20 | 4 | 0.0676 | -0.4958 | 0.0157 | 2 | 0.0666 | -0.6086 |  | 2 | 0.0579 | -0.7036 |  |
| 21 | 2 | 0.0755 | -0.5279 |  | 3 | 0.0754 | -0.6401 | 0.0102 | 4 | 0.0658 | -0.7249 | 0.0209 |
| 22 | 3 | 0.0826 | -0.5515 | 0.0083 | 2 | 0.0825 | -0.6679 |  | 4 | 0.0763 | -0.7632 | 0.0223 |
| 23 | 2 | 0.0917 | -0.5893 |  | 3 | 0.0915 | -0.6930 | 0.0123 | 2 | 0.0842 | -0.7880 |  |
| 24 | 4 | 0.0990 | -0.6114 | 0.0126 | 4 | 0.1007 | -0.7168 | 0.0200 | 3 | 0.0933 | -0.7958 | 0.0095 |
| 25 | 4 | 0.1084 | -0.6423 | 0.0100 | 3 | 0.1089 | -0.7452 | 0.0200 | 3 | 0.1014 | -0.8240 | 0.0219 |
| 26 | 3 | 0.1172 | -0.6612 | 0.0104 | 5 | 0.1163 | -0.7470 | 0.0112 | 2 | 0.1091 | -0.8532 |  |
| 27 | 4 | 0.1250 | -0.6819 | 0.0042 | 6 | 0.1251 | -0.7710 | 0.0107 | 3 | 0.1155 | -0.8688 | 0.0261 |
| 28 | 6 | 0.1336 | -0.6956 | 0.0075 | 4 | 0.1336 | -0.7923 | 0.0142 | 2 | 0.1256 | -0.8981 |  |
| 29 | 3 | 0.1422 | -0.7105 | 0.0043 | 4 | 0.1404 | -0.8028 | 0.0151 | 5 | 0.1325 | -0.8875 | 0.0220 |
| 30 | 4 | 0.1508 | -0.7287 | 0.0085 | 4 | 0.1502 | -0.8203 | 0.0095 | 5 | 0.1413 | -0.9062 | 0.0187 |
| 31 | 2 | 0.1603 | -0.7473 |  | 5 | 0.1586 | -0.8329 | 0.0162 | 3 | 0.1507 | -0.9320 | 0.0185 |
| 32 | 5 | 0.1670 | -0.7578 | 0.0054 | 5 | 0.1669 | -0.8411 | 0.0092 | 4 | 0.1590 | -0.9340 | 0.0182 |
| 33 | 3 | 0.1747 | -0.7734 | 0.0057 | 4 | 0.1746 | -0.8596 | 0.0136 | 5 | 0.1670 | -0.9478 | 0.0191 |
| 34 | 3 | 0.1832 | -0.7805 | 0.0078 | 4 | 0.1836 | -0.8686 | 0.0184 | 4 | 0.1762 | -0.9609 | 0.0168 |
| 35 | 4 | 0.1922 | -0.7945 | 0.0050 | 3 | 0.1927 | -0.8729 | 0.0128 | 3 | 0.1840 | -0.9664 | 0.0189 |
| 36 | 2 | 0.2003 | -0.8025 |  | 3 | 0.2004 | -0.8881 | 0.0162 | 3 | 0.1915 | -0.9858 | 0.0222 |
| 37 | 4 | 0.2082 | -0.8103 | 0.0060 | 3 | 0.2088 | -0.8846 | 0.0129 | 2 | 0.1982 | -0.9862 |  |
| 38 | 3 | 0.2174 | -0.8194 | 0.0046 | 3 | 0.2161 | -0.9031 | 0.0187 | 3 | 0.2075 | -1.0042 | 0.0239 |
| 39 | 3 | 0.2244 | -0.8249 | 0.0069 | 4 | 0.2250 | -0.9025 | 0.0189 | 2 | 0.2172 | -1.0350 |  |
| 40 | 4 | 0.2334 | -0.8298 | 0.0058 | 3 | 0.2340 | -0.9013 | 0.0144 | 3 | 0.2256 | -1.0070 | 0.0308 |
| 41 | 2 | 0.2409 | -0.8327 |  | 3 | 0.2410 | -0.9094 | 0.0159 | 2 | 0.2322 | -1.0367 |  |
| 42 | 3 | 0.2482 | -0.8426 | 0.0124 | 3 | 0.2498 | -0.9126 | 0.0107 | 3 | 0.2410 | -1.0308 | 0.0314 |
| 43 | 1 | 0.2612 | -0.8292 |  | 4 | 0.2661 | -0.9238 | 0.0079 | 2 | 0.2490 | -1.0193 |  |
| 44 | 3 | 0.2655 | -0.8384 | 0.0078 | 3 | 0.2749 | -0.9228 | 0.0077 | 1 | 0.2548 | -1.0009 |  |
| 45 | 4 | 0.2744 | -0.8366 | 0.0086 | 3 | 0.2821 | -0.9179 | 0.0122 | 3 | 0.2666 | -1.0320 | 0.0130 |
| 46 | 3 | 0.2832 | -0.8297 | 0.0058 | 4 | 0.2909 | -0.9176 | 0.0070 | 3 | 0.2737 | -1.0203 | 0.0080 |
| 47 | 3 | 0.2904 | -0.8309 | 0.0092 | 3 | 0.2998 | -0.9048 | 0.0071 | 4 | 0.2826 | -1.0242 | 0.0120 |
| 48 | 4 | 0.2993 | -0.8219 | 0.0068 | 3 | 0.3071 | -0.9051 | 0.0066 | 3 | 0.2915 | -1.0219 | 0.0111 |
| 49 | 3 | 0.3085 | -0.8116 | 0.0077 | 3 | 0.3159 | -0.8908 | 0.0068 | 3 | 0.2987 | -1.0110 | 0.0106 |
| 50 | 2 | 0.3155 | -0.8087 |  | 4 | 0.3247 | -0.8856 | 0.0085 | 3 | 0.3088 | -1.0015 | 0.0081 |
| 51 | 3 | 0.3243 | -0.7950 | 0.0073 | 3 | 0.3335 | -0.8778 | 0.0062 | 2 | 0.3188 | -0.9976 |  |
| 52 | 4 | 0.3331 | -0.7858 | 0.0119 | 3 | 0.3409 | -0.8678 | 0.0040 | 3 | 0.3251 | -0.9899 | 0.0164 |
| 53 | 3 | 0.3419 | -0.7754 | 0.0071 | 3 | 0.3502 | -0.8567 | 0.0049 | 3 | 0.3325 | -0.9877 | 0.0152 |
| 54 | 3 | 0.3493 | -0.7609 | 0.0106 | 3 | 0.3587 | -0.8466 | 0.0049 | 4 | 0.3414 | -0.9775 | 0.0184 |
| 55 | 3 | 0.3571 | -0.7531 | 0.0089 | 3 | 0.3670 | -0.8309 | 0.0135 | 3 | 0.3502 | -0.9590 | 0.0186 |
| 56 | 2 | 0.3671 | -0.7303 |  | 4 | 0.3763 | -0.8126 | 0.0085 | 1 | 0.3580 | -0.9388 |  |
| 57 | 4 | 0.3747 | -0.7211 | 0.0087 | 2 | 0.3840 | -0.7907 |  | 2 | 0.3642 | -0.9442 |  |
| 58 | 2 | 0.3846 | -0.7058 |  | 3 | 0.3923 | -0.7763 | 0.0154 | 2 | 0.3755 | -0.9210 |  |
| 59 | 3 | 0.3908 | -0.6865 | 0.0084 | 3 | 0.4006 | -0.7617 | 0.0066 | 3 | 0.3838 | -0.9067 | 0.0129 |
| 60 | 4 | 0.4000 | -0.6694 | 0.0118 | 2 | 0.4092 | -0.7365 |  | 3 | 0.3922 | -0.8920 | 0.0180 |
| 61 | 3 | 0.4090 | -0.6481 | 0.0088 | 1 | 0.4137 | -0.7292 |  | 3 | 0.3991 | -0.8857 | 0.0224 |
| 62 | 1 | 0.4271 | -0.5919 |  | 1 | 0.4330 | -0.6610 |  | 3 | 0.4092 | -0.8578 | 0.0178 |
| 63 | 2 | 0.4340 | -0.5762 |  | 3 | 0.4422 | -0.6381 | 0.0095 | 3 | 0.4164 | -0.8322 | 0.0162 |
| 64 | 2 | 0.4426 | -0.5488 |  | 3 | 0.4507 | -0.6064 | 0.0162 | 1 | 0.4233 | -0.8034 |  |
| 65 | 4 | 0.4502 | -0.5287 | 0.0104 | 2 | 0.4569 | -0.5943 |  | 2 | 0.4424 | -0.7446 |  |
| 66 | 3 | 0.4593 | -0.5043 | 0.0084 | 4 | 0.4673 | -0.5712 | 0.0083 | 3 | 0.4496 | -0.7276 | 0.0089 |
| 67 | 3 | 0.4668 | -0.4831 | 0.0068 | 3 | 0.4762 | -0.5524 | 0.0102 | 4 | 0.4587 | -0.7045 | 0.0102 |
| 68 | 3 | 0.4746 | -0.4687 | 0.0083 | 4 | 0.4835 | -0.5422 |  | 3 | 0.4677 | -0.6818 | 0.0037 |
| 69 | 3 | 0.4826 | -0.4539 | 0.0059 | 2 | 0.4910 | -0.5329 |  | 3 | 0.4751 | -0.6648 | 0.0049 |
| 70 | 3 | 0.4912 | -0.4478 | 0.0056 | 2 | 0.4991 | -0.5274 |  | 2 | 0.4815 | -0.6585 |  |
| 71 | 1 | 0.5000 | -0.4478 |  | 4 | 0.5090 | -0.5369 | 0.0080 | 2 | 0.4926 | -0.6532 |  |
| 72 | 4 | 0.5074 | -0.4478 | 0.0042 | 2 | 0.5165 | -0.5456 |  | 2 | 0.5007 | -0.6466 |  |

| No. | No. of Points | Phase | Δb | P.E. | No. of Points | Phase | Δv | P.E. | No. of Points | Phase | Δr | P.E. |
|---|---|---|---|---|---|---|---|---|---|---|---|---|
| 73 | 4 | 0.5175 | -0.4635 | 0.0039 | 4 | 0.5242 | -0.5619 | 0.0071 | 2 | 0.5081 | -0.6499 | |
| 74 | 2 | 0.5251 | -0.4793 | | 3 | 0.5331 | -0.5823 | 0.0104 | 4 | 0.5157 | -0.6573 | 0.0051 |
| 75 | 3 | 0.5315 | -0.4948 | 0.0070 | 2 | 0.5418 | -0.6120 | | 4 | 0.5258 | -0.6740 | 0.0069 |
| 76 | 3 | 0.5416 | -0.5200 | 0.0106 | 3 | 0.5487 | -0.6284 | 0.0134 | 3 | 0.5346 | -0.7002 | 0.0182 |
| 77 | 3 | 0.5488 | -0.5351 | 0.0053 | 3 | 0.5592 | -0.6648 | 0.0124 | 3 | 0.5416 | -0.7175 | 0.0194 |
| 78 | 2 | 0.5586 | -0.5601 | | 3 | 0.5656 | -0.6901 | 0.0140 | 3 | 0.5499 | -0.7374 | 0.0117 |
| 79 | 1 | 0.5663 | -0.5822 | | 4 | 0.5742 | -0.7085 | 0.0041 | 2 | 0.5567 | -0.7704 | |
| 80 | 3 | 0.5756 | -0.6118 | 0.0171 | 3 | 0.5831 | -0.7325 | 0.0085 | 1 | 0.5696 | -0.8067 | |
| 81 | 3 | 0.5841 | -0.6363 | 0.0088 | 3 | 0.5915 | -0.7481 | 0.0185 | 1 | 0.5747 | -0.8207 | |
| 82 | 3 | 0.5919 | -0.6595 | 0.0048 | 2 | 0.6001 | -0.7656 | | 4 | 0.5836 | -0.8306 | 0.0084 |
| 83 | 4 | 0.6010 | -0.6769 | 0.0053 | 4 | 0.6077 | -0.7805 | 0.0090 | 2 | 0.5911 | -0.8549 | |
| 84 | 2 | 0.6086 | -0.6947 | | 4 | 0.6178 | -0.7965 | 0.0098 | 4 | 0.5992 | -0.8725 | 0.0103 |
| 85 | 3 | 0.6171 | -0.7099 | 0.0038 | 2 | 0.6253 | -0.8066 | | 4 | 0.6093 | -0.8893 | 0.0096 |
| 86 | 3 | 0.6252 | -0.7222 | 0.0053 | 2 | 0.6326 | -0.8159 | | 2 | 0.6169 | -0.8996 | |
| 87 | 1 | 0.6335 | -0.7339 | | 3 | 0.6419 | -0.8346 | 0.0084 | 3 | 0.6254 | -0.9097 | 0.0074 |
| 88 | 2 | 0.6409 | -0.7470 | | 1 | 0.6497 | -0.8506 | | 2 | 0.6342 | -0.9328 | |
| 89 | 2 | 0.6507 | -0.7694 | | 2 | 0.6573 | -0.8610 | | 1 | 0.6419 | -0.9410 | |
| 90 | 1 | 0.6581 | -0.7769 | | 3 | 0.6670 | -0.8625 | 0.0132 | 2 | 0.6488 | -0.9409 | |
| 91 | 4 | 0.6665 | -0.7900 | 0.0044 | 2 | 0.6732 | -0.8733 | | 2 | 0.6589 | -0.9558 | |
| 92 | 2 | 0.6758 | -0.8041 | | 4 | 0.6833 | -0.8756 | 0.0104 | 2 | 0.6673 | -0.9621 | |
| 93 | 3 | 0.6828 | -0.8070 | 0.0087 | 2 | 0.6909 | -0.8815 | | 4 | 0.6748 | -0.9740 | 0.0079 |
| 94 | 4 | 0.6920 | -0.8147 | 0.0020 | 2 | 0.7014 | -0.8893 | | 3 | 0.6838 | -0.9826 | 0.0107 |
| 95 | 3 | 0.7012 | -0.8255 | 0.0043 | 4 | 0.7089 | -0.8925 | 0.0134 | 3 | 0.6911 | -0.9878 | 0.0024 |
| 96 | 3 | 0.7083 | -0.8258 | 0.0101 | 2 | 0.7180 | -0.9040 | | 4 | 0.7005 | -0.9955 | 0.0122 |
| 97 | 2 | 0.7164 | -0.8319 | | 3 | 0.7258 | -0.8944 | 0.0257 | 3 | 0.7094 | -0.9955 | 0.0023 |
| 98 | 4 | 0.7254 | -0.8375 | 0.0155 | 3 | 0.7332 | -0.9006 | 0.0085 | 2 | 0.7155 | -0.9981 | |
| 99 | 3 | 0.7343 | -0.8397 | 0.0006 | 3 | 0.7410 | -0.8913 | 0.0084 | 3 | 0.7248 | -1.0045 | 0.0132 |
| 100 | 3 | 0.7418 | -0.8370 | 0.0098 | 4 | 0.7500 | -0.8886 | 0.0180 | 3 | 0.7348 | -1.0040 | 0.0047 |
| 101 | 3 | 0.7497 | -0.8328 | 0.0112 | 2 | 0.7585 | -0.9038 | | 2 | 0.7413 | -1.0009 | |
| 102 | 3 | 0.7570 | -0.8364 | 0.0169 | 3 | 0.7664 | -0.8912 | 0.0199 | 2 | 0.7503 | -0.9982 | 0.0026 |
| 103 | 3 | 0.7653 | -0.8379 | 0.0114 | 2 | 0.7734 | -0.8895 | | 2 | 0.7567 | -0.9974 | |
| 104 | 1 | 0.7743 | -0.8347 | | 2 | 0.7834 | -0.8796 | | 3 | 0.7653 | -0.9978 | 0.0045 |
| 105 | 2 | 0.7818 | -0.8227 | | 1 | 0.7909 | -0.8714 | | 3 | 0.7736 | -0.9854 | 0.0015 |
| 106 | 1 | 0.7893 | -0.8183 | | 2 | 0.8009 | -0.8621 | | 1 | 0.7825 | -0.9841 | |
| 107 | 2 | 0.7993 | -0.8129 | | 1 | 0.8084 | -0.8518 | | 2 | 0.7901 | -0.9712 | |
| 108 | 2 | 0.8094 | -0.7972 | | 2 | 0.8161 | -0.8309 | | 1 | 0.8000 | -0.9591 | |
| 109 | 1 | 0.8170 | -0.7887 | | 2 | 0.8260 | -0.8171 | | 2 | 0.8076 | -0.9436 | |
| 110 | 2 | 0.8244 | -0.7735 | | 1 | 0.8334 | -0.7898 | | 2 | 0.8177 | -0.9337 | |
| 111 | 2 | 0.8343 | -0.7625 | | 2 | 0.8408 | -0.7883 | | 1 | 0.8251 | -0.9223 | |
| 112 | 1 | 0.8417 | -0.7435 | | 1 | 0.8575 | -0.7770 | | 2 | 0.8325 | -0.9066 | |
| 113 | 2 | 0.8584 | -0.7178 | | 2 | 0.8650 | -0.7677 | | 2 | 0.8425 | -0.8942 | |
| 114 | 1 | 0.8659 | -0.7098 | | 2 | 0.8750 | -0.7452 | | 1 | 0.8514 | -0.8791 | |
| 115 | 2 | 0.8734 | -0.6924 | | 1 | 0.8825 | -0.7277 | | 1 | 0.8591 | -0.8766 | |
| 116 | 2 | 0.8834 | -0.6692 | | 2 | 0.8901 | -0.7067 | | 2 | 0.8666 | -0.8641 | |
| 117 | 1 | 0.8910 | -0.6385 | | 2 | 0.9002 | -0.6752 | | 1 | 0.8740 | -0.8398 | |
| 118 | 2 | 0.8985 | -0.6265 | | 1 | 0.9085 | -0.6507 | | 2 | 0.8817 | -0.8351 | |
| 119 | 2 | 0.9094 | -0.5880 | | 2 | 0.9159 | -0.6154 | | 2 | 0.8917 | -0.8039 | |
| 120 | 1 | 0.9167 | -0.5775 | | 2 | 0.9258 | -0.5872 | | 1 | 0.8992 | -0.7951 | |
| 121 | 2 | 0.9242 | -0.5322 | | 2 | 0.9863 | -0.3960 | | 2 | 0.9072 | -0.7562 | |
| 122 | 2 | 0.9847 | -0.2734 | | 2 | 0.9923 | -0.4034 | | 2 | 0.9175 | -0.7204 | |
| 123 | 3 | 0.9921 | -0.2647 | 0.0043 | 4 | 0.9998 | -0.4023 | 0.0072 | 1 | 0.9249 | -0.7061 | |
| 124 | 2 | 0.9992 | -0.2593 | 0.0022 | 3 | 1.0088 | -0.4090 | 0.0073 | 1 | 0.9299 | -0.6667 | |
| 125 | 4 | 1.0082 | -0.2711 | 0.0034 | 3 | 1.0159 | -0.4284 | 0.0075 | 4 | 0.9910 | -0.5341 | 0.0211 |
| 126 | 3 | 1.0172 | -0.2896 | 0.0093 | 3 | 1.0249 | -0.4515 | 0.0071 | 3 | 1.0004 | -0.5367 | 0.0121 |
| 127 | 3 | 1.0243 | -0.3145 | 0.0168 | 3 | 1.0341 | -0.4844 | 0.0147 | 3 | 1.0075 | -0.5341 | 0.0146 |
| 128 | 4 | 1.0335 | -0.3462 | 0.0139 | 2 | 1.0403 | -0.5028 | | 4 | 1.0165 | -0.5553 | 0.0143 |
| 129 | 1 | 1.0457 | -0.3997 | | 3 | 1.0499 | -0.5526 | 0.0195 | 3 | 1.0255 | -0.5829 | 0.0170 |
| 130 | 3 | 1.0509 | -0.4194 | 0.0118 | 4 | 1.0589 | -0.5861 | 0.0131 | 3 | 1.0326 | -0.5967 | 0.0226 |
| 131 | 3 | 1.0584 | -0.4600 | 0.0187 | 2 | 1.0666 | -0.6086 | | 3 | 1.0409 | -0.6396 | 0.0280 |
| 132 | 4 | 1.0676 | -0.4958 | 0.0157 | 3 | 1.0754 | -0.6401 | 0.0102 | 3 | 1.0515 | -0.6925 | 0.0348 |
| 133 | 2 | 1.0755 | -0.5279 | | 2 | 1.0825 | -0.6679 | | 2 | 1.0579 | -0.7036 | |
| 134 | 3 | 1.0826 | -0.5515 | 0.0083 | 3 | 1.0915 | -0.6930 | 0.0123 | 4 | 1.0658 | -0.7249 | 0.0209 |
| 135 | 2 | 1.0917 | -0.5893 | | 4 | 1.1007 | -0.7168 | 0.0200 | 4 | 1.0763 | -0.7632 | 0.0223 |
| 136 | 4 | 1.0990 | -0.6114 | 0.0126 | 3 | 1.1089 | -0.7452 | 0.0200 | 2 | 1.0842 | -0.7880 | |
| 137 | 4 | 1.1084 | -0.6423 | 0.0100 | 5 | 1.1163 | -0.7470 | 0.0112 | 3 | 1.0933 | -0.7958 | 0.0095 |
| 138 | 3 | 1.1172 | -0.6612 | 0.0104 | 6 | 1.1251 | -0.7710 | 0.0107 | 3 | 1.1014 | -0.8240 | 0.0219 |
| 139 | 4 | 1.1250 | -0.6819 | 0.0042 | 4 | 1.1336 | -0.7923 | 0.0142 | 2 | 1.1091 | -0.8532 | |
| 140 | 6 | 1.1336 | -0.6956 | 0.0075 | 4 | 1.1404 | -0.8028 | 0.0151 | 3 | 1.1155 | -0.8688 | 0.0261 |
| 141 | 3 | 1.1422 | -0.7105 | 0.0043 | 2 | 1.1475 | -0.8127 | | 2 | 1.1256 | -0.8981 | |
| 142 | 1 | 1.1469 | -0.7221 | | | | | | 5 | 1.1325 | -0.8875 | 0.0220 |
| 143 | | | | | | | | | 5 | 1.1413 | -0.9062 | 0.0187 |
| 144 | | | | | | | | | 1 | 1.1481 | -0.9403 | |